\documentclass[twocolumn]{aastex62}
\graphicspath{{./}{}}
\shorttitle{Fiber Dithering}
\shortauthors{Schlafly et al.}
\usepackage{textcomp}
\usepackage{hyperref}
\usepackage{amsmath}
\usepackage{amsfonts}
\newcommand{\degree}{\ensuremath{^\circ}}
\newcommand{\mum}{\ensuremath{\mu}m}

\begin{document}

\title{Measuring Fiber Positioning Accuracy and Throughput with Fiber Dithering for the Dark Energy Spectroscopic Instrument}


\author[0000-0002-3569-7421]{E.~F.~Schlafly}
\affiliation{Space Telescope Science Institute, 3700 San Martin Drive, Baltimore, MD 21218, USA}

\author{D.~Schlegel}
\affiliation{Lawrence Berkeley National Laboratory, 1 Cyclotron Road, Berkeley, CA 94720, USA}

\author[0000-0001-5537-4710]{S.~BenZvi}
\affiliation{Department of Physics \& Astronomy, University of Rochester, 206 Bausch and Lomb Hall, P.O. Box 270171, Rochester, NY 14627-0171, USA}

\author[0000-0001-5999-7923]{A.~Raichoor}
\affiliation{Lawrence Berkeley National Laboratory, 1 Cyclotron Road, Berkeley, CA 94720, USA}

\author[0000-0002-2890-3725]{J.~E.~Forero-Romero}
\affiliation{Departamento de F\'isica, Universidad de los Andes, Cra. 1 No. 18A-10, Edificio Ip, CP 111711, Bogot\'a, Colombia}
\affiliation{Observatorio Astron\'omico, Universidad de los Andes, Cra. 1 No. 18A-10, Edificio H, CP 111711 Bogot\'a, Colombia}

\author{J.~Aguilar}
\affiliation{Lawrence Berkeley National Laboratory, 1 Cyclotron Road, Berkeley, CA 94720, USA}

\author[0000-0001-6098-7247]{S.~Ahlen}
\affiliation{Physics Dept., Boston University, 590 Commonwealth Avenue, Boston, MA 02215, USA}

\author[0000-0003-4162-6619]{S.~Bailey}
\affiliation{Lawrence Berkeley National Laboratory, 1 Cyclotron Road, Berkeley, CA 94720, USA}

\author[0000-0002-9964-1005]{A.~Bault}
\affiliation{Department of Physics and Astronomy, University of California, Irvine, 92697, USA}

\author{D.~Brooks}
\affiliation{Department of Physics \& Astronomy, University College London, Gower Street, London, WC1E 6BT, UK}

\author{T.~Claybaugh}
\affiliation{Lawrence Berkeley National Laboratory, 1 Cyclotron Road, Berkeley, CA 94720, USA}

\author{K.~Dawson}
\affiliation{Department of Physics and Astronomy, The University of Utah, 115 South 1400 East, Salt Lake City, UT 84112, USA}

\author[0000-0002-1769-1640]{A.~de la Macorra}
\affiliation{Instituto de F\'{\i}sica, Universidad Nacional Aut\'{o}noma de M\'{e}xico,  Cd. de M\'{e}xico  C.P. 04510,  M\'{e}xico}

\author[0000-0002-4928-4003]{Arjun~Dey}
\affiliation{NSF's NOIRLab, 950 N. Cherry Ave., Tucson, AZ 85719, USA}

\author{P.~Doel}
\affiliation{Department of Physics \& Astronomy, University College London, Gower Street, London, WC1E 6BT, UK}

\author{E.~Gaztañaga}
\affiliation{Institut d'Estudis Espacials de Catalunya (IEEC), 08034 Barcelona, Spain}
\affiliation{Institute of Cosmology \& Gravitation, University of Portsmouth, Dennis Sciama Building, Portsmouth, PO1 3FX, UK}
\affiliation{Institute of Space Sciences, ICE-CSIC, Campus UAB, Carrer de Can Magrans s/n, 08913 Bellaterra, Barcelona, Spain}

\author[0000-0003-3142-233X]{S.~Gontcho A Gontcho}
\affiliation{Lawrence Berkeley National Laboratory, 1 Cyclotron Road, Berkeley, CA 94720, USA}

\author{J.~Guy}
\affiliation{Lawrence Berkeley National Laboratory, 1 Cyclotron Road, Berkeley, CA 94720, USA}

\author[0000-0003-1197-0902]{C.~Hahn}
\affiliation{Department of Astrophysical Sciences, Princeton University, Princeton NJ 08544, USA}

\author{K.~Honscheid}
\affiliation{Center for Cosmology and AstroParticle Physics, The Ohio State University, 191 West Woodruff Avenue, Columbus, OH 43210, USA}
\affiliation{Department of Physics, The Ohio State University, 191 West Woodruff Avenue, Columbus, OH 43210, USA}

\author[0000-0001-8528-3473]{J.~Jimenez}
\affiliation{Institut de F\'{i}sica d’Altes Energies (IFAE), The Barcelona Institute of Science and Technology, Campus UAB, 08193 Bellaterra Barcelona, Spain}

\author[0000-0003-4207-7420]{S.~Kent}
\affiliation{Department of Astronomy and Astrophysics, University of Chicago, 5640 South Ellis Avenue, Chicago, IL 60637, USA}
\affiliation{Fermi National Accelerator Laboratory, PO Box 500, Batavia, IL 60510, USA}

\author[0000-0002-8828-5463]{D.~Kirkby}
\affiliation{Department of Physics and Astronomy, University of California, Irvine, 92697, USA}

\author[0000-0003-3510-7134]{T.~Kisner}
\affiliation{Lawrence Berkeley National Laboratory, 1 Cyclotron Road, Berkeley, CA 94720, USA}

\author[0000-0001-6356-7424]{A.~Kremin}
\affiliation{Lawrence Berkeley National Laboratory, 1 Cyclotron Road, Berkeley, CA 94720, USA}

\author{A.~Lambert}
\affiliation{Lawrence Berkeley National Laboratory, 1 Cyclotron Road, Berkeley, CA 94720, USA}

\author[0000-0003-1838-8528]{M.~Landriau}
\affiliation{Lawrence Berkeley National Laboratory, 1 Cyclotron Road, Berkeley, CA 94720, USA}

\author[0000-0003-1887-1018]{Michael~E.~Levi}
\affiliation{Lawrence Berkeley National Laboratory, 1 Cyclotron Road, Berkeley, CA 94720, USA}

\author[0000-0003-4962-8934]{M.~Manera}
\affiliation{Departament de F\'{i}sica, Serra H\'{u}nter, Universitat Aut\`{o}noma de Barcelona, 08193 Bellaterra (Barcelona), Spain}
\affiliation{Institut de F\'{i}sica d’Altes Energies (IFAE), The Barcelona Institute of Science and Technology, Campus UAB, 08193 Bellaterra Barcelona, Spain}

\author[0000-0002-4279-4182]{P.~Martini}
\affiliation{Center for Cosmology and AstroParticle Physics, The Ohio State University, 191 West Woodruff Avenue, Columbus, OH 43210, USA}
\affiliation{Department of Astronomy, The Ohio State University, 4055 McPherson Laboratory, 140 W 18th Avenue, Columbus, OH 43210, USA}

\author[0000-0002-1125-7384]{A.~Meisner}
\affiliation{NSF's NOIRLab, 950 N. Cherry Ave., Tucson, AZ 85719, USA}

\author{R.~Miquel}
\affiliation{Instituci\'{o} Catalana de Recerca i Estudis Avan\c{c}ats, Passeig de Llu\'{\i}s Companys, 23, 08010 Barcelona, Spain}
\affiliation{Institut de F\'{i}sica d’Altes Energies (IFAE), The Barcelona Institute of Science and Technology, Campus UAB, 08193 Bellaterra Barcelona, Spain}

\author[0000-0002-2733-4559]{J.~Moustakas}
\affiliation{Department of Physics and Astronomy, Siena College, 515 Loudon Road, Loudonville, NY 12211, USA}

\author{A.~D.~Myers}
\affiliation{Department of Physics \& Astronomy, University  of Wyoming, 1000 E. University, Dept.~3905, Laramie, WY 82071, USA}

\author[0000-0001-6590-8122]{J.~Nie}
\affiliation{National Astronomical Observatories, Chinese Academy of Sciences, A20 Datun Rd., Chaoyang District, Beijing, 100012, P.R. China}

\author[0000-0003-3188-784X]{N.~Palanque-Delabrouille}
\affiliation{IRFU, CEA, Universit\'{e} Paris-Saclay, F-91191 Gif-sur-Yvette, France}
\affiliation{Lawrence Berkeley National Laboratory, 1 Cyclotron Road, Berkeley, CA 94720, USA}

\author[0000-0002-0644-5727]{W.~J.~Percival}
\affiliation{Department of Physics and Astronomy, University of Waterloo, 200 University Ave W, Waterloo, ON N2L 3G1, Canada}
\affiliation{Perimeter Institute for Theoretical Physics, 31 Caroline St. North, Waterloo, ON N2L 2Y5, Canada}
\affiliation{Waterloo Centre for Astrophysics, University of Waterloo, 200 University Ave W, Waterloo, ON N2L 3G1, Canada}

\author{C.~Poppett}
\affiliation{Lawrence Berkeley National Laboratory, 1 Cyclotron Road, Berkeley, CA 94720, USA}
\affiliation{Space Sciences Laboratory, University of California, Berkeley, 7 Gauss Way, Berkeley, CA  94720, USA}
\affiliation{University of California, Berkeley, 110 Sproul Hall \#5800 Berkeley, CA 94720, USA}

\author[0000-0001-7145-8674]{F.~Prada}
\affiliation{Instituto de Astrof\'{i}sica de Andaluc\'{i}a (CSIC), Glorieta de la Astronom\'{i}a, s/n, E-18008 Granada, Spain}

\author{D.~Rabinowitz}
\affiliation{Physics Department, Yale University, P.O. Box 208120, New Haven, CT 06511, USA}

\author[0000-0001-5589-7116]{M.~Rezaie}
\affiliation{Department of Physics, Kansas State University, 116 Cardwell Hall, Manhattan, KS 66506, USA}

\author{G.~Rossi}
\affiliation{Department of Physics and Astronomy, Sejong University, Seoul, 143-747, Korea}

\author[0000-0002-9646-8198]{E.~Sanchez}
\affiliation{CIEMAT, Avenida Complutense 40, E-28040 Madrid, Spain}

\author{M.~Schubnell}
\affiliation{Department of Physics, University of Michigan, Ann Arbor, MI 48109, USA}

\author[0000-0003-3449-8583]{R.~Sharples}
\affiliation{Centre for Advanced Instrumentation, Department of Physics, Durham University, South Road, Durham DH1 3LE, UK}
\affiliation{Institute for Computational Cosmology, Department of Physics, Durham University, South Road, Durham DH1 3LE, UK}

\author[0000-0002-3461-0320]{J.~Silber}
\affiliation{Lawrence Berkeley National Laboratory, 1 Cyclotron Road, Berkeley, CA 94720, USA}

\author[0000-0003-1704-0781]{G.~Tarl\'{e}}
\affiliation{Department of Physics, University of Michigan, Ann Arbor, MI 48109, USA}

\author{B.~A.~Weaver}
\affiliation{NSF's NOIRLab, 950 N. Cherry Ave., Tucson, AZ 85719, USA}

\author[0000-0002-4135-0977]{Z.~Zhou}
\affiliation{National Astronomical Observatories, Chinese Academy of Sciences, A20 Datun Rd., Chaoyang District, Beijing, 100012, P.R. China}

\author[0000-0002-6684-3997]{H.~Zou}
\affiliation{National Astronomical Observatories, Chinese Academy of Sciences, A20 Datun Rd., Chaoyang District, Beijing, 100012, P.R. China}

\collaboration{(DESI Collaboration)}

\begin{abstract}
Highly multiplexed, fiber-fed spectroscopy is enabling surveys of millions of stars and galaxies.  The performance of these surveys depends on accurately positioning fibers in the focal plane to capture target light.  We describe a technique to measure the positioning accuracy of fibers by dithering fibers slightly around their ideal locations.  This approach also enables measurement of the total system throughput and point spread function delivered to the focal plane.  We then apply this technique to observations from the Dark Energy Survey Instrument (DESI), and demonstrate that DESI positions fibers to within 0.08\arcsec\ of their targets (5\%\ of a fiber diameter) and achieves a system throughput within about 5\% of expectations.
\end{abstract}



\section{Introduction} 
\label{sec:intro}

Highly multiplexed fiber-fed spectroscopic systems are enabling a number of major current and upcoming astronomical surveys, like the Dark Energy Spectroscopic Instrument \citep[DESI,][]{Levi:2013}, the Prime Focus Spectrograph \citep{Takada:2014}, the SDSS-V \citep{Kollmeier:2017}, 4MOST \citep{deJong:2019}, LAMOST \citep{Cui:2012}, and 2dF \citep{Colless:2001}.  These systems depend on the ability to position fibers precisely in the focal plane where target light is brought into focus.  For example, for DESI, where fibers are 107~\mum\ in diameter, positioning errors of only 10~\micron\ lead to flux losses of 2\% and decrease the survey speed by 4\% \citep{desi:2016a, desi:2016b}.  Systematic errors in fiber positioning as a function of location in the focal plane can also lead to spatial trends in the redshift accuracy and success rate of the main DESI survey, complicating downstream cosmological analyses \citep[e.g.][]{Krolewski:2024, Yu:2024}.  Accurate positioning of fibers in the focal plane is then critical to the success of these systems.

The approach taken to positioning fibers depends on the details of the instrument.  However, broadly, imaging cameras are used to position the telescope at the intended location on the sky and to guide the telescope during the observation.  With the telescope's location fixed on the sky, next the mapping of the focal plane to the sky and the locations of the fibers in the focal plane must be determined.  The former is often limited by imperfect knowledge of the optics of the system, while the latter can be limited by imperfect metrology and uncertainty in the measurement of fibers' positions.  In the case of DESI, a special camera (the Fiber View Camera, or FVC) can image the fibers to improve the measurement of the fibers' positions in the focal plane, though this introduces additional uncertainties stemming from imperfect knowledge of the optical system of that camera.  Analysis of the optical system and the system metrology can lead to good predictions for the on-sky locations of each fiber through this approach, but it is important to be able to assess the accuracy of these predictions.

Assessing the fiber positioning accuracy of spectroscopic systems can be a significant challenge.  Often the primary observable is the amount of light entering the spectrograph, which depends on the point spread function (PSF) delivered to the fibers.  This observed flux is sensitive to the magnitude of the positioning error, but not its direction.  Moreover, the PSF delivered to the fiber is usually unknown.  Often estimates of the PSF are available from guide cameras elsewhere in the system, but the PSF at the location of a given fiber may be different than the PSF delivered to the guider.  This can lead to an ambiguous situation where more or less light is entering the spectrograph than expected, but where the origin of the discrepancy is unclear.  Since the amount of source light entering the spectrograph is perhaps the most important contributor to the speed at which a survey proceeds, resolving this discrepancy is critically important.

Here we present a technique that overcomes these difficulties by directly measuring fiber positioning accuracy and therefore system throughput.  We achieve this by intentionally displacing fibers away from bright stellar targets with accurate astrometry by known amounts.  This approach is conceptually related to moving fibers around target stars and looking for the flux to ``peak up'' when the fiber is correctly positioned \citep[e.g.][]{Hill:1988}.  By intentionally observing off of the nominal locations we can observe how the flux changes with position to determine where to best position the fiber.

For highly multiplexed systems, intentionally displacing fibers provides additional benefits, however.  The key idea is that in many systems, the dominant source of positioning error is systematic and fixed over a sequence of exposures, while time-variable parameters describing the PSF and system throughput apply to all of the fibers in each exposure.  By displacing different fibers by different amounts, the system PSF can be measured, and the positioning offsets can be inferred from the knowledge of the PSF and the amount of flux obtained at each offset position.  Knowledge of the PSF and displacements in turn allows the system throughput to be measured without ambiguity by comparing the observed flux with the amount of flux expected from imaging data.  This technique can be applied to any highly multiplexed fiber-fed spectroscopic system.

This approach is an approximation.  For example, the PSF varies across the field from fiber to fiber.  Turbulence in the volume of air between the FVC and the corrector leads to positioning errors that are not constant among fibers or between exposures.  Uncorrected field rotation blurs the PSF on the edge of the focal plane more than in the center of the focal plane.  But if effects like these are well enough controlled relative to the systematic positioning errors this approximation can be productive.

We apply this fiber-dithering technique to special engineering observations from DESI, demonstrating that DESI positions fibers with an accuracy of 0.08\arcsec\ and delivers a total throughput within about 5\% of expectations.  This accuracy was achieved via the construction of detailed maps of the distortions in the DESI optics, which we measured using analysis of these observations.

This paper consists of the following sections.  In \textsection\ref{sec:method}, we lay out the method we use to measure fiber positioning accuracy and throughput via fiber dithering.  In \textsection\ref{sec:sims}, we simulate a variety of dither strategies and show how the recovered displacement accuracy depends on the approach used.  In \textsection\ref{sec:desi}, we apply this technique to observations from DESI, and discuss our results in \textsection\ref{sec:results}.  Finally, we conclude in \textsection\ref{sec:conclusion}.  The code and data used to produce the tables and figures in this paper are available at \url{https://zenodo.org/doi/10.5281/zenodo.10693684}.

\section{Method}
\label{sec:method}

We intend to use a series of observations in which the fibers have been intentionally offset by known amounts to measure the systematic fiber offsets from true target positions, the per-exposure throughput, and the per-exposure PSF.  These observations give the spectra of a number of target stars over a series of exposures, where the known offset is varied for each source from exposure to exposure.  We construct a forward model of these observations as a function of the parameters of interest---the fiber offsets, point spread function, transparency, and any guide errors.  We then optimize the parameters of the model to determine the best fit parameters.  

\begin{figure}
    \centering
    \includegraphics[width=\columnwidth]{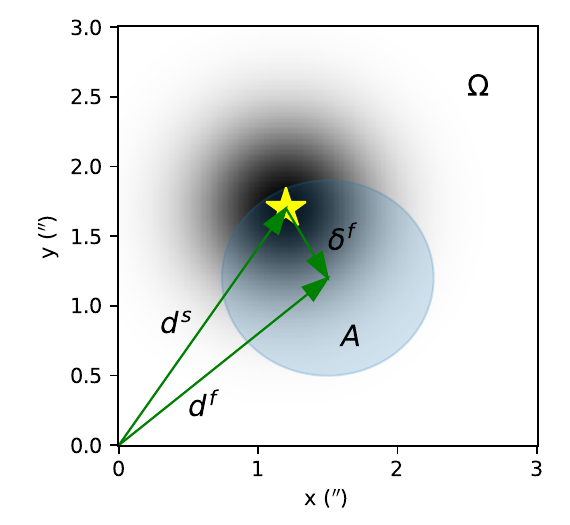}
    \caption{Schematic diagram of the basic model of this work.  A star emits light over a broad PSF (background grayscale).  Light entering the fiber (blue shaded region, described by $A$) gets collected by the spectrograph.  The ellipticity of the fiber $A$ on the sky is appropriate for a DESI fiber 400 mm from the center of the focal plane, and is hardly visible.  Each dither of a fiber positioner places the fiber center $d^f$ at a different position with respect to the star at $d_s$, capturing different amounts of light.}
    \label{fig:schematic}
\end{figure}

The basic observable we need to model is the flux entering the spectrograph as a function of the offset between the fiber and the source.  The flux $F$ at wavelength $\lambda$ recorded in a spectrograph from fiber $i$ is given by
\begin{equation}
\label{eq:full}
F_i(\lambda) = \int d\Omega \, S(\lambda) T(\lambda) A_i(\Omega-d^f_i) P(\Omega-d^s_i, \lambda) \, .
\end{equation}
The flux is the integral over the area of the fiber projected on the sky. For each position $\Omega$ on the sky, $A_i$ determines if the sky location $\Omega$ enters the fiber when centered at $d_f$, and $P$ is the value of the PSF, where $P(\Omega - d_s)$ is the value at $\Omega$ given that the star is centered at $d_s$.  The function $S(\lambda)$ gives the spectral energy distribution of the sources as a function of wavelength, and $T$ describes the system throughput.  The function $A_i$ is 1 when a portion of sky is mapped into the fiber aperture and 0 otherwise, and depends on $i$ to account for the potentially varying plate scale across the focal plane; see \textsection\ref{subsec:desifiberacceptance} for details of our implementation.  We are here implicitly assuming that the PSF $P$ is constant over the focal plane; a more general model would allow this to vary.  Figure~\ref{fig:schematic} schematically shows the geometry of the problem.  The important positions and positional offsets we end up using are tabulated in Table~\ref{tab:possymbols}.

We can simplify the integrand in Equation~\ref{eq:full} to remove the wavelength dependence by defining a particular wavelength band $b$ and integrating over all wavelengths in the band, giving
\begin{equation}
    \int d\lambda \, T_b(\lambda) S(\lambda) P(\Omega-d^s, \lambda) = a_b P_{b, S}(\Omega-d^s) \, .
\end{equation}
In a particular band $b$, the target star has an observed flux $a_b$ and an effective PSF $P_b$.  As long as the band is reasonably narrow, we can ignore the dependence of $P_{b, S}$ on the source spectrum $S$, and so we drop this subscript in the following discussion.  Changing variables, the observed flux in the band is given by 
\begin{equation}
    F_{i, b} = a_b \int d\Omega \, P_b(\Omega) A_i(\Omega-(d^f_i-d^s_i)) \, .
\end{equation}
Note that here the meaning of $\Omega$ has changed, and in earlier equations corresponds to $\Omega - d^s_i$.  This equation gives the flux of an observed star in an observation.  It contains three unknowns---the flux of the star and the two coordinates of $d^f_i - d^s_i$---and one observable, the flux $F$.

\begin{deluxetable}{ll}
    \tablecaption{\label{tab:possymbols} Glossary of symbols for positions and offsets}
\tablehead{
 \colhead{Symbol} & \colhead{Description}}
\startdata
$d^s_i$ & location of source \\
$d^f_i$ & location of fiber $i$ \\
$d^f_{0,i}$ & location of fiber absent intentional dither  \\
$\delta^f_{0,i}$ & systematic positioning error \\
$\delta^f_{e,i}$ & intentional fiber offset from nominal location \\
$\delta^t_e$ & telescope guide offset; one per exposure
\enddata
\tablenotetext{}{The various different positions and offsets used in this paper.  All locations are on sky.
}
\end{deluxetable}

Though the number of unknowns exceeds the number of measurements, we can constrain these parameters if we make a sequence of $N_\mathrm{exp}$ exposures $e$ observing the same set of stars, where we dither the fibers by a set of offsets $\delta^f_{e,i}$ so that $d^f_i = d^f_{0,i} + \delta^f_{e,i}$.  The total flux of the star remains the same, and the $\delta^f_{e,i}$ are known, provided that the fibers can be precisely dithered, so the main uncertainty is in $d^f_{0,i}$, the true location of the fiber on the sky.  In this case, we have $N_\mathrm{exp}$ observations of fluxes but retain the three unknowns (two components of $d^f_{0,i}$ and the flux $a_b$), making the problem well posed.

Moreover, if we have $N_\mathrm{fiber}$ fibers in each exposure, we can also solve for exposure-wide parameters such as the overall motion of the telescope on the sky relative to the desired location, changes in the atmospheric transparency, and parameters describing the PSF.  This leads to the model:
\begin{equation}
F_{b, e, i} = a_{b, i} T_e \int d\Omega \, P_b(\Omega, \theta_e) A_f(\Omega-(\delta^f_{0,i}+\delta^f_{e, i}+\delta^t_e)) \, ,
\label{eq:fullmodel}
\end{equation}
where $F_{b, e, i}$ is the observed flux in exposure $e$ and fiber $i$, $T_e$ is the throughput of exposure $e$, $\delta^t_e$ is the overall pointing offset of the telescope in exposure $e$, and $\theta_e$ are any parameters describing the PSF in exposure $e$.  We have also additionally introduced $\delta^f_{0,i} = d^f_{0,i} - d^s_i$, the systematic positioning offset of each fiber that we intend to measure.

For the PSF, we adopt a Moffat profile with an unknown FWHM, position angle, ellipticity, and power law index.  With this parameterization of the PSF, the total number of parameters is $7N_\mathrm{exp}$ (throughput $T$, guide errors $\delta^t_e$ in $x$ and $y$, and 4 PSF shape parameters) plus $3N_\mathrm{fiber}$ (the star fluxes $a_{b,i}$ and the fiber positioning offsets $\delta^f_{0, i}$ in $x$ and $y$), with $N_\mathrm{fiber}N_\mathrm{exp}$ measurements.  For typical highly multiplexed systems, $N_\mathrm{fiber} \gg 7$, so these systems become well constrained when $N_{\mathrm{exp}} > 3$.  In principle only four exposures are needed to measure the position of each fiber and the system throughput, though in practice for DESI we use $N_\mathrm{exp} = 13$.

Equation~\ref{eq:fullmodel} gives the flux entering a particular fiber as a function of the model parameters.  To derive these model parameters from a set of observed fluxes, we define a Gaussian likelihood function
\begin{equation}
\label{eq:loglike}
    \log L_b = -\frac{1}{2} \sum_{e=1}^{N_\mathrm{exp}} \sum_{i=1}^{N_\mathrm{fiber}} \frac{(F_{\mathrm{obs},b,e,i} - F_{\mathrm{model},b,e,i})^2}{\sigma_{b,e,i}^2} \, .
\end{equation}
The uncertainties $\sigma_{b,e,i}$ must contain at least the Poisson noise in the fluxes and may contain additional systematic contributions; our implementation for DESI is described in \textsection\ref{subsec:quantities}.  We solve this model using Levenberg-Marquardt optimization, minimizing the negative log likelihood or equivalently $\chi^2$.  The integral over the the fiber aperture in Equation~\ref{eq:fullmodel} makes the model non-analytic, requiring the evaluation of numerical derivatives during minimization.  The model has many parameters, so these numerical derivatives are expensive to compute if one naively re-evaluates the full likelihood when computing the derivative with respect to each parameter.  However, exposure parameters (e.g. $T$, FWHM) only affect fluxes measured in that exposure, and fiber parameters (e.g., the fiber positioning errors) only affect fluxes in that fiber.  Accordingly, given the exposure parameters, each fiber can be solved separately and in parallel, and given the fiber parameters, each exposure can be solved separately and in parallel.  This allows the optimization to be dramatically accelerated by alternately solving for the exposure and fiber parameters, separating a many-parameter problem into a series of few-parameter problems.  Solutions take a few minutes using 60 cores and ten iterations alternating between the per-exposure and per-fiber parameters.  We start by assuming that the fibers are well centered and have fluxes consistent with the image fluxes, and iteratively improve the per-exposure and per-fiber parameters until reaching convergence, typically running 10 iterations.

We define three bands, $B$, $R$, and $Z$, and solve for all of the parameters in each band independently.  The bands are defined by integrating the DESI system throughput over the wavelengths given in Table~\ref{tab:bandpasses}.  The wavelength limits are chosen to lie entirely within a DESI spectrograph and are similar to the DECam $g$, $r$, and $z$ bands, facilitating comparison of imaging fluxes with spectroscopic fluxes for determining total system throughput.  Solutions for different bands allow a variety of useful tests.  For example, the DESI optical design leads to small changes in the center of light in the focal plane as a function of wavelength.  The differences of derived fiber offsets at different wavelengths can be compared with expectations from the optical design (see \textsection\ref{subsec:chromatic}).  Comparison of dither results in different bands also allows us to test the consistency of the fiber positioning offsets with different data sets.  For spectroscopic systems like DESI, we could choose any set of wavelengths; the choices of Table~\ref{tab:bandpasses} are merely convenient.

\begin{deluxetable}{cccc}
    \tablecaption{\label{tab:bandpasses}}
\tablehead{
 \colhead{camera} & \colhead{imaging band} & \colhead{blue limit (nm)} & \colhead{red limit (nm)}}
\startdata
B & g & 400 & 550 \\
R & r & 565 & 712 \\
Z & z & 850 & 990 \\
\enddata
\tablenotetext{}{The bandpasses used for the spectroscopic fluxes analyzed in this work.  We integrate over the DESI spectra between these wavelength limits.
}
\end{deluxetable}



\section{Simulations}
\label{sec:sims}

We run simulations to study the ability of fiber dithering to recover the fluxes of stars, the positions of fibers, and the guider offsets, throughputs, and PSFs of exposures.  The performance of the algorithm depends importantly on the pattern of fiber dithers $\delta^f_{e,i}$.  If they are too small, they do not constrain the PSF of the instrument well, while if they are too large, little light is collected and the results are dominated by background.

In the simulations we model fiber offsets $\delta^f_0$ as independent random variables.  For DESI, this is a poor approximation: most sources of systematic error in fiber position are highly correlated in the focal plane.  The chief cause of systematic error is imperfections in the mapping between the location of fibers in back-lit images of the focal plane and their true location in the focal plane, which are highly correlated between fibers \citep{Kent:2023}.  However, because the modeling approach does not try to use any information about correlated offsets in positioner locations, this limitation of our simulations does not invalidate our results.  On the contrary, it conservatively bounds our performance: better performance would be possible if we included in the model terms that favored correlations in fiber offsets.

We run a simple set of simulations where the seeing is modeled as constant in arcseconds on the sky across the focal plane.  The seeing varies from exposure to exposure following a Gaussian distribution with a mean of 1.1\arcsec\ and a standard deviation of 0.2\arcsec.  We assume that 40\% of the flux entering a fiber reaches the spectrograph, with a standard deviation of 1\%.  Fibers have systematic positioning errors across all exposures in the dither sequence described by a uniform distribution from $-0.1\arcsec$ to $+0.1\arcsec$.  Telescope guide errors are drawn from a Gaussian $\mathcal{N}(0\arcsec, 0.1\arcsec)$.  We assume stars are chosen with magnitudes such that the number of photons that would enter the spectrograph absent fiber acceptance losses is uniformly distributed between 5,000 and 10,000.  We simulate 5000 observed stars on each exposure, matching DESI's 5000 fibers, but note that this is generous---real dither sequences end up with closer to 4000 fibers on bright stars, due to allocating some fibers to sky measurement ($\sim 500$) and other fibers being non-functional or not having access to a sufficiently bright star. With these input conditions, we then simulate how different dithering schemes produce different delivered fluxes across exposures, and how that leads to varying constraints on the fiber positioning errors.

\begin{figure*}
    \centering
    \includegraphics[width=\textwidth]{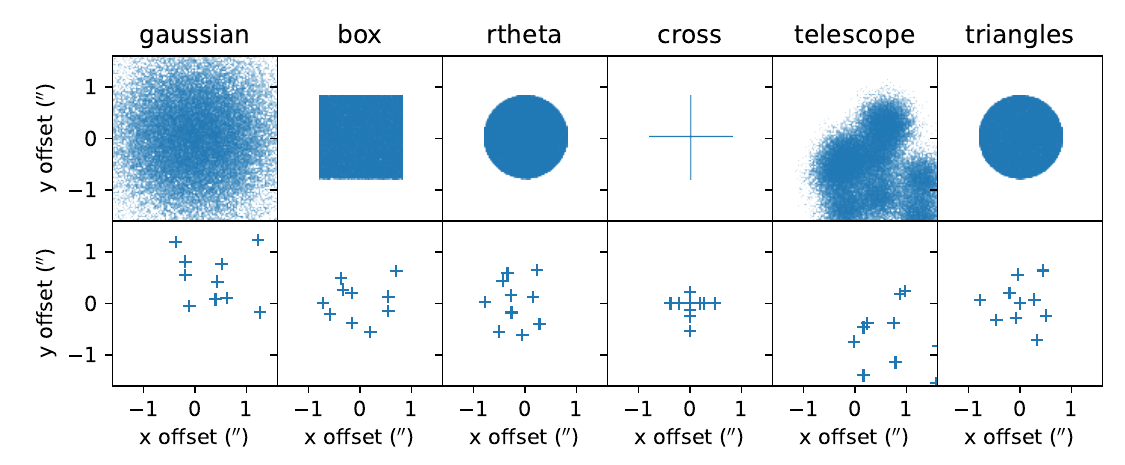}
    \caption{Dither schemes simulated in order to assess how fiber positioning measurements depend on the dithering scheme.  The different columns correspond to different patterns of dithers.  The top row shows the set of dithers for the full 5000 positioners simulated, while the bottom row shows one particular positioner, chosen at random among the 5000.}
    \label{fig:ditherschemes}
\end{figure*}

We generate ten dithered exposures using the above model, considering a number of different possible fiber dither patterns $\delta^f_{e,i}$.  Each pattern is parameterized by a scale factor $\sigma$ which controls how large we should make the dithers.  The simulated patterns are:
\begin{enumerate}
    \item A Gaussian set of dithers; fibers are randomly dithered by $\mathcal{N}(0, \sigma)$ in each coordinate in each exposure.
    \item A ``box'' set of dithers: fibers are randomly dithered by $U(-\sigma, +\sigma)$ in each coordinate.
    \item A ``disc'' set of dithers: fiber dithers are chosen from a disc centered at 0 with a radius of $\sigma$.
    \item A ``cross'' set of dithers.  Either the $x$ or $y$ direction is chosen at random for each exposure and each fiber.  Fibers are uniformly dithered by $U(-\sigma, \sigma)$ in that direction.
    \item A ``telescope'' set of dithers.  Fibers are dithered a single time in advance, and then the telescope boresight is moved from exposure to exposure by a fixed amount.  This can be practical, for example, when the time needed to dither fibers is substantial.  Fiber dithers are drawn from a $\mathcal{N}(0, 0.5\sigma)$ distribution in each coordinate, and telescope dithers are drawn from a $\mathcal{N}(0, \sigma)$ distribution in the two telescope coordinates.
    \item A ``triangle'' set of dithers; ten fibers are allocated in the following way: one fiber is directly on target; the remaining nine fibers are allocated three each to three rings, separated by 120\degree\ around the ring.  The radii of the rings are chosen such that the final points are uniformly distributed by area, and each fiber gets one ring in an inner, one ring in a middle, and one ring in an outer zone.  The idea here is to get the overall distribution of the ``disc'' dithers, but making sure that every fiber gets good azimuthal and radial coverage individually.
\end{enumerate}
Figure~\ref{fig:ditherschemes} illustrates the different schemes, showing the distribution of dither offsets for all 5,000 fibers and for one particular fiber.  These schemes are meant to cover a broad range of possible ideas about how one might want to dither fibers around targets, but we have not formally attempted to figure out the optimal assignment.  We ultimately find only modest differences in performance between these schemes.  We note that the ``telescope'' dither scheme can be used on systems where dithering fibers between exposures is expensive (e.g., for plug plate systems used in the Sloan Digital Sky Survey, \citealt{York:2000}).

After choosing a dithering scheme, we apply Equation~\ref{eq:fullmodel} to generate simulated data corresponding to a dither sequence, and model the data following \textsection\ref{sec:method} to derive the best fit parameters.  We compare the true parameters entering the simulation with the fits.  We do these comparisons for each dither scheme over a range of dither scales and tabulate the results in Table~\ref{tab:simresults}.  We find formal uncertainties in positioner offsets as good as $0.006\arcsec$, though in practice systematic effects not present in the simulation prevent us from reaching this precision.  Given DESI's $1.5\arcsec$ fiber aperture, there are diminishing returns to be had for positioning accuracy better than about $0.1\arcsec$ \citep{desiinstrumentation:2022}.   

\begin{deluxetable*}{cccccccc}
\tablecaption{Accuracy of parameter recovery in simulations\label{tab:simresults}}
\tablehead{
 pattern, scale & \colhead{$\delta^f_{0, x}$ (\arcsec)}& \colhead{$\delta^f_{0, y}$ (\arcsec)} & \colhead{mag} & \colhead{FWHM (\arcsec)} & \colhead{$\delta^t_x$ (\arcsec)} & \colhead{$\delta^t_y$ (\arcsec)} & \colhead{$T$}}
\startdata
gaussian, 0.2\arcsec & $ 0.0089$ & $ 0.0090$ & $ 0.0073$ & $ 0.0013$ & $ 0.0004$ & $ 0.0007$ & $ 0.0017$ \\
gaussian, 0.4\arcsec & $ 0.0066$ & $ 0.0065$ & $ 0.0081$ & $ 0.0006$ & $ 0.0003$ & $ 0.0004$ & $ 0.0007$ \\
gaussian, 0.8\arcsec & $ 0.0077$ & $ 0.0077$ & $ 0.0128$ & $ 0.0005$ & $ 0.0002$ & $ 0.0005$ & $ 0.0007$ \\
gaussian, 1.6\arcsec & $ 0.0247$ & $ 0.0237$ & $ 0.0530$ & $ 0.0015$ & $ 0.0007$ & $ 0.0010$ & $ 0.0009$ \\
gaussian, 3.2\arcsec & $ 0.5634$ & $ 0.5506$ & $ 1.5132$ & $ 0.0058$ & $ 0.0056$ & $ 0.0040$ & $ 0.0123$ \\
\hline
box, 0.2\arcsec & $ 0.0125$ & $ 0.0119$ & $ 0.0073$ & $ 0.0052$ & $ 0.0022$ & $ 0.0039$ & $ 0.0062$ \\
box, 0.4\arcsec & $ 0.0080$ & $ 0.0078$ & $ 0.0073$ & $ 0.0029$ & $ 0.0004$ & $ 0.0007$ & $ 0.0031$ \\
box, 0.8\arcsec & $ 0.0060$ & $ 0.0058$ & $ 0.0090$ & $ 0.0008$ & $ 0.0003$ & $ 0.0004$ & $ 0.0006$ \\
box, 1.6\arcsec & $ 0.0103$ & $ 0.0099$ & $ 0.0199$ & $ 0.0007$ & $ 0.0003$ & $ 0.0003$ & $ 0.0009$ \\
box, 3.2\arcsec & $ 0.1861$ & $ 0.1906$ & $ 0.5902$ & $ 0.0035$ & $ 0.0029$ & $ 0.0029$ & $ 0.0062$ \\
\hline
rtheta, 0.2\arcsec & $ 0.0137$ & $ 0.0133$ & $ 0.0072$ & $ 0.0159$ & $ 0.0045$ & $ 0.0067$ & $ 0.0166$ \\
rtheta, 0.4\arcsec & $ 0.0086$ & $ 0.0086$ & $ 0.0072$ & $ 0.0028$ & $ 0.0005$ & $ 0.0008$ & $ 0.0032$ \\
rtheta, 0.8\arcsec & $ 0.0059$ & $ 0.0060$ & $ 0.0083$ & $ 0.0010$ & $ 0.0002$ & $ 0.0003$ & $ 0.0009$ \\
rtheta, 1.6\arcsec & $ 0.0082$ & $ 0.0080$ & $ 0.0157$ & $ 0.0007$ & $ 0.0003$ & $ 0.0004$ & $ 0.0008$ \\
rtheta, 3.2\arcsec & $ 0.0876$ & $ 0.0809$ & $ 0.2272$ & $ 0.0034$ & $ 0.0021$ & $ 0.0021$ & $ 0.0028$ \\
\hline
cross, 0.2\arcsec & $ 0.0167$ & $ 0.0162$ & $ 0.0077$ & $ 0.0191$ & $ 0.0085$ & $ 0.0095$ & $ 0.0210$ \\
cross, 0.4\arcsec & $ 0.0108$ & $ 0.0105$ & $ 0.0073$ & $ 0.0035$ & $ 0.0017$ & $ 0.0016$ & $ 0.0063$ \\
cross, 0.8\arcsec & $ 0.0069$ & $ 0.0069$ & $ 0.0075$ & $ 0.0004$ & $ 0.0004$ & $ 0.0003$ & $ 0.0006$ \\
cross, 1.6\arcsec & $ 0.0075$ & $ 0.0076$ & $ 0.0103$ & $ 0.0005$ & $ 0.0004$ & $ 0.0003$ & $ 0.0008$ \\
cross, 3.2\arcsec & $ 0.0148$ & $ 0.0144$ & $ 0.0203$ & $ 0.0007$ & $ 0.0006$ & $ 0.0006$ & $ 0.0014$ \\
\hline
telescope, 0.2\arcsec & $ 0.0105$ & $ 0.0119$ & $ 0.0139$ & $ 0.0429$ & $ 0.0257$ & $ 0.0266$ & $ 0.0421$ \\
telescope, 0.4\arcsec & $ 0.0073$ & $ 0.0084$ & $ 0.0139$ & $ 0.0215$ & $ 0.0134$ & $ 0.0172$ & $ 0.0138$ \\
telescope, 0.8\arcsec & $ 0.0094$ & $ 0.0113$ & $ 0.0215$ & $ 0.0080$ & $ 0.0081$ & $ 0.0121$ & $ 0.0174$ \\
telescope, 1.6\arcsec & $ 0.0239$ & $ 0.0266$ & $ 0.0679$ & $ 0.0598$ & $ 0.0158$ & $ 0.0828$ & $ 0.3475$ \\
telescope, 3.2\arcsec & $ 0.4550$ & $ 0.5276$ & $ 1.6098$ & $ 0.4415$ & $ 0.0300$ & $ 0.1390$ & $ 0.1829$ \\
\hline
triangles, 0.2\arcsec & $ 0.0130$ & $ 0.0128$ & $ 0.0070$ & $ 0.0165$ & $ 0.0044$ & $ 0.0039$ & $ 0.0196$ \\
triangles, 0.4\arcsec & $ 0.0082$ & $ 0.0081$ & $ 0.0065$ & $ 0.0023$ & $ 0.0008$ & $ 0.0004$ & $ 0.0031$ \\
triangles, 0.8\arcsec & $ 0.0057$ & $ 0.0057$ & $ 0.0072$ & $ 0.0009$ & $ 0.0003$ & $ 0.0003$ & $ 0.0011$ \\
triangles, 1.6\arcsec & $ 0.0073$ & $ 0.0073$ & $ 0.0110$ & $ 0.0007$ & $ 0.0004$ & $ 0.0003$ & $ 0.0007$ \\
triangles, 3.2\arcsec & $ 0.0226$ & $ 0.0233$ & $ 0.0206$ & $ 0.0014$ & $ 0.0013$ & $ 0.0015$ & $ 0.0024$ \\
\enddata
\tablenotetext{}{The root-mean-square difference between simulated model parameters and those recovered from fitting simulated data, for the fiber offsets in $x$ and $y$ ($\delta^f_0$), the total stellar magnitudes, the telescope FWHM, the telescope pointing offsets in $x$ and $y$ ($\delta^t$), and the transparency.  
}
\end{deluxetable*}

We draw three general conclusions from Table~\ref{tab:simresults}.  First, there are enough photons to measure fiber positioning offsets at $<0.1\arcsec$ uncertainty.  Second, for 1\arcsec\ seeing, a dither pattern scale of $\sim 0.8\arcsec$ gives good performance.  This roughly matches expectations.  Dithers much smaller than the point spread function do not allow the shape of the PSF to be well measured, and do not probe the part of the PSF where flux varies rapidly with offset, and therefore fail to provide useful information in a dither analysis.  Meanwhile, very large dithers result in little flux entering a fiber and therefore are not very sensitive to the the position of the fiber.  The two factors balance when the dither scale is about the scale of the fiber-aperture convolved seeing.  For DESI, with a median seeing of $\sim 1\arcsec$ and a fiber aperture size of $1.5\arcsec$, that corresponds to a Gaussian $\sigma$ of about $0.8\arcsec$.  Third, performance is not very sensitive to the particular dither scheme or scale; varying the scale by a factor of two only changes performance by a factor of about 50\%, and similar performance is obtained for the hand-tuned ``triangles'' scheme and the Gaussian scheme.

\section{Dither Observations with DESI}
\label{sec:desi}

During DESI's commissioning period \citep[see][]{desisv, desiedr}, we applied this dither modeling approach to determine fiber positioning errors and help improve DESI's positioning performance.  Between January 1, 2020 and March 14, 2020 we improved the typical positioning errors from $\sim10\arcsec$ to $\sim0.1\arcsec$.  The dither analysis played an important role in this process.  At the start of commissioning, few fibers recorded meaningful amounts of light, and it was not completely clear which stars were illuminating the few fibers that in fact recorded large fluxes.  Eventually, dither information revealed that the hexapod was significantly rotated relative to our expectations, and after correcting the field rotation we were able to place all of the fibers near their targets.  Subsequent improvements focused on reducing the amount of distortion in the FVC lens \citep{Baltay:2019}, which we achieved by replacing the FVC lens in early Feburary 2020 after fiber dither modeling showed significant, roughly arcsecond-scale positioning errors.  With the new lens in place, the positioning residuals dropped to about $0.4\arcsec$, which we accounted for by including a static, low-order, dither-derived correction in the positioning process.  This correction enabled the $\sim0.1\arcsec$ positioning precision required for the start of the DESI main survey.  We discuss the details of the dither observations below.

\subsection{How DESI positions fibers}
The basic technique of this work to determining fiber positioning accuracy can be applied to any large multi-object spectrograph.  However, to interpret the results of this technique applied to DESI it is useful to understand more about the DESI system and how DESI position fibers.  For more details about DESI, see reviews of the survey and instrument in \citet{Levi:2013, desi:2016a, desi:2016b, desiinstrumentation:2022}, and for more information in particular about the positioning of fibers, see \citet{Kent:2023}.

The DESI focal plane is at the prime focus of the Mayall telescope, a 4~m equatorial telescope with a 3.2\degree\ field of view \citep{Miller:2023}.  DESI's 5,000 fibers fill this focal plane and are controlled by robotic positioners, which place the fibers anywhere within a 1.4\arcmin\ radius of the central location of each positioner \citep{Silber:2023}.  The DESI focal plane is divided into 10 nearly identical petals with 500 fibers each.  Each petal is connected to a three-armed spectrograph, where the $B$ arm covers 360~nm to 580~nm, the $R$ arm 570~nm to 760~nm, and the $Z$ arm 760~nm to 980~nm.  When requested to make a new observation, DESI first performs a ``blind'' move, followed by a ``correction'' move.  The blind move uses our expectation for the world coordinate system (WCS) mapping to the focal plane to move the DESI positioners close to their intended locations.  The ``correction'' move improves on these locations, using the field WCS following acquisition images from the guide focus array cameras (GFAs) and feedback from the FVC about where each of the fibers landed after the blind move.

The FVC resides near the hole in the center of the primary mirror, and looks through the corrector onto the focal plane \citep{desiinstrumentation:2022}.  DESI's fibers can be back lit from LED strips on the shutters in the spectrographs.  When back-lit, the FVC can observe the apparent locations of each of the fibers relative to the locations of the fiducials on the GFAs \citep{Kent:2023, Silber:2023}.  These positions are taken as effectively the locations of the fibers in the tangent plane of the sky, and are converted to on-sky positions using astrometric information from the GFAs.  Following the blind move, fibers are back-lit, imaged by the FVC, and their locations on the sky are measured.  DESI then makes a correction move, adjusting the locations of the fibers from their measured locations to their intended final positions.  This move is able to correct for imperfections in the calibration of the positioners (i.e., the positioners were requested to go to the wrong place), errors in the positioners' motion (i.e., the positioners didn't end up exactly where requested), and errors in the assumed world-coordinate system (e.g., the system had a slightly different focus or scale factor than presumed for the blind move).  Following the correction move, the FVC takes an additional back-lit image of the fibers, which we take as the final best estimate of where each positioner was located for an exposure.

This system allows for accurate measurements of small fiber offsets even when overall telescope guiding errors and absolute fiber positions are poorly known.  Small fiber offset measurements depend only on the local plate scale and the ability to measure the light centroids from the fibers in the FVC images, which are relatively easily measured.  Additionally, small fiber moves are more accurate than large ones, mostly due to their reduced dependence on the positioners' calibration parameters.  This feature allows dither analyses, which depend on accurate measurements of the small fiber dithers ($\delta^f_{e,i}$) to proceed early in commissioning when many aspects of the system are not fully understood.

\subsection{Target selection}
\label{subsec:targetselection}

The dither analysis requires that positioners be dithered around bright, isolated stars.  Bright targets give better uncertainties in the delivered fluxes, and isolated stars allow simplified modeling of the flux as a function of location.  We have two different kinds of selection serving this broad goal: one for regions with Legacy Survey imaging \citep{Dey:2019, Zou:2017}, and one for regions where only measurements from Gaia are available \citep{Gaia:2016}.

In detail, our dither target selection for areas with Legacy Survey imaging includes the following cuts:
\begin{itemize}
    \item The Legacy Survey imaging type must be `PSF'.
    \item The source must be in Gaia DR2.
    \item There must be little contamination from neigbors---\texttt{fracflux\_g}, \texttt{fracflux\_r}, \texttt{fracflux\_z} must all be less than 0.002 \citep{Dey:2019}.
    \item The Gaia $G$ and $RP$ magnitudes must be fainter than 11.5 to avoid saturating the detector.
    \item The Gaia astrometric fits must be good---the astrometric excess noise must be less than 1.
    \item The Gaia proper motions must be finite.
    \item Gaia must not consider the source a ``duplicated source''.
\end{itemize}
Broadly, these cuts select all bright, isolated point sources that DESI can observe without saturating.  We set priorities on the sources during fiber assignment so that bright sources are assigned before faint ones, so that the brightest of the selected stars are assigned fibers.

When no Legacy Survey imaging is available, the Gaia portions of these cuts translate over directly, but we assess whether the source is isolated using a different approach.  Additionally, to avoid adding too many sources to the DESI targeting catalogs, we reduce the density of targets somewhat, adding the following cuts:
\begin{itemize}
    \item All sources within 7\arcsec\ must be more than 100$\times$ fainter than this source.
    \item Gaia $G < 19$ mag if $|b| < 20\degree$, otherwise $G < 20$ mag.
\end{itemize}
Because we have Legacy Survey imaging available at essentially all high Galactic latitude fields observable by DESI, we only use the Gaia-only selection at low Galactic latitudes.  At these latitudes there are many stars available and so faint stars can be trimmed without any loss to the dither program.

All targets are selected through the DESI target selection system of \citet{Myers:2023}.

\subsection{Observations}
We chose to use the ``Gaussian'' dither scheme from \textsection\ref{sec:method}, with a scale of $0.7\arcsec$.  This was close to the best scale for the ``Gaussian'' method, biased a bit high to try to be a bit more robust to positioners that might be very poorly positioned.  The ``Gaussian'' scheme was chosen more for simplicity than due to the detailed simulation results.

We designed a sequence of 13 ``tiles'' on each dither field using the DESI \texttt{fiberassign} software.  Here, a ``tile'' refers to a particular assignment of fibers to target positions.  We designed an initial ``on-target'' tile where fibers were placed directly on targets, followed by 12 dithered tiles.  Targets were taken from the on-target tile and target lists were restricted to only targets in the on-target tile for subsequent tiles.  The on-target tiles are useful for validating throughput measurements but are not very useful in the context of the overall dither analysis; since all fibers are on target, the PSF is poorly measured, and since the fiber acceptance is maximized when on target, small positioning errors have only small effects on the delivered flux.

These 13 tiles were then observed with DESI, passing through the DESI Instrument Control System and downstream spectroscopic reduction pipeline \citep{Guy:2023} as ordinary tiles.  We flag dither tiles in the reduction pipeline as only requiring analysis through sky subtraction, however; subsequent steps like flux calibration are not needed for dithered observations.

We have observed a number of dither sequences with DESI.  During early commissioning starting in January 2020, on many nights we took multiple dither sequences as we tried to understand initial fiber positioning problems.  Early in the main survey, starting May 14, 2021, we usually took a dither sequence around full moon to verify continued good fiber positioning performance.  More recently we have taken dither sequences only occasionally during engineering time and following significant changes to the instrument.  For example, we repeated dither observations after the major DESI summer shutdowns in 2021 and 2022 \citep{Schlafly:2023}.

This work focuses on nine dither sequences taken in good conditions (photometric over most of the sequence, all petals operating, seeing better than $1.5\arcsec$), taken between September 2021 and April 2023.  Observational details of these dither sequences are given in Table~\ref{tab:dithersequences}.  We note that the first and second dither sequences in Table~\ref{tab:dithersequences} were taken in the same part of the sky and are nearly identical.  Additionally the third, fifth, and sixth sequences all used the same sequence of tiles, and are different only in the particular observational conditions during those observations.  Finally, Table~\ref{tab:dithersequences} also includes a tenth dither sequence from October 2023, which was taken after the higher-order distortion correction of this work was incorporated into the pipeline.  We do not include the results of this sequence in most of the analysis here, but nevertheless include it in Table~\ref{tab:dithersequences} and Figure~\ref{fig:ditherquiver} to demonstrate the success of the correction. 

\begin{deluxetable*}{lrrrlcccc}
    \tablecaption{\label{tab:dithersequences}}
\tablehead{
 \colhead{date} & \colhead{first tile id} & \colhead{$\alpha$} & \colhead{$\delta$} & \colhead{exposure ids} & alt & \colhead{seeing} & off & off$^\prime$}
\startdata
2021--09--16\tablenotemark{a} & 82130 & 336.0\degree & 30.0\degree & 100469--100481 & 87\degree & 0.9\arcsec & 0.16\arcsec & 0.09\arcsec\\
2021--10--19 & 82269 & 336.0\degree & 30.0\degree & 105197--105202 & 75\degree & 0.6\arcsec & 0.14\arcsec & 0.07\arcsec \\
2021--10--23 & 82282 & 1.6\degree & 31.2\degree & 105777--105798 & 85\degree & 1.2\arcsec & 0.13\arcsec & 0.07\arcsec \\
2021--10--24 & 82308 & 52.5\degree & 37.5\degree & 105900--105912 & 42\degree & 1.3\arcsec & 0.14\arcsec & 0.09\arcsec \\
2021--12--16 & 82282 & 1.6\degree & 31.2\degree & 114321--114333 & 73\degree & 0.9\arcsec & 0.15\arcsec & 0.09\arcsec \\
2021--12--19 & 82282 & 1.6\degree & 31.2\degree & 114724--114736 & 82\degree & 0.9\arcsec & 0.14\arcsec & 0.07\arcsec \\
2022--05--18 & 82731 & 279.0\degree & 50.0\degree & 135607--135620 & 72\degree & 0.7\arcsec & 0.14\arcsec & 0.07\arcsec \\
2022--09--13 & 82360 & 2.6\degree & 54.0\degree & 142240--142252 & 67\degree & 0.8\arcsec & 0.14\arcsec & 0.09\arcsec \\
2023--04--11 & 82705 & 220.2\degree & 47.8\degree & 175918--175930 & 73\degree & 0.7\arcsec & 0.15\arcsec & 0.08\arcsec \\
\hline
2023--10--30\tablenotemark{b} & 82360 & 2.6\degree & 54.0\degree & 202772--202784 & 63\degree & 1.7\arcsec & 0.08\arcsec & ---
\enddata
\tablenotetext{}{The dither observations discussed in this work.  The ``off'' (``off$^\prime$'') column gives $\sqrt{\langle \delta_x^2 + \delta_y^2 \rangle }$ for the derived positioning errors $\delta^f_0$ before (after) correction for a static positioning offset pattern.  Right ascensions ($\alpha$) and declinations ($\delta$) are given for the center of each tile.  Dither observations cover a range of airmass but consistently deliver positioning offsets of between 0.13\arcsec--0.16\arcsec\ before correction and $\approx0.08\arcsec$ afterward.
}
\tablenotetext{a}{The final three exposures of this sequence had transparencies ranging from 0.9 to 0.1 as clouds rolled in.}
\tablenotetext{b}{This sequence was taken after the higher order distortion correction described in this work was incorporated into the pipeline, and has significantly lower positional offsets than the other sequences.  It is excluded from most of the analysis of this paper but demonstrates the success of the correction.} 
\end{deluxetable*}

We observe each tile in a dither sequence for three minutes.  Typical per-tile overhead is 100~s, so the total time needed to run a dither sequence is about one hour.  We obtain adequate signal-to-noise ratios on most targets in significantly less than three minutes---thirty or sixty seconds still provide plenty of signal for these bright stars---but we observe for the full three minutes to allow the telescope tracking and guiding to fully engage.  As much as possible, we want the dither observations to match normal observations, so that the positioning performance in dithered observations will well represent the positioning performance in real observations.

\subsection{The amount of light entering a DESI fiber}
\label{subsec:desifiberacceptance}

\begin{figure}
    \centering
    \includegraphics[width=\columnwidth]{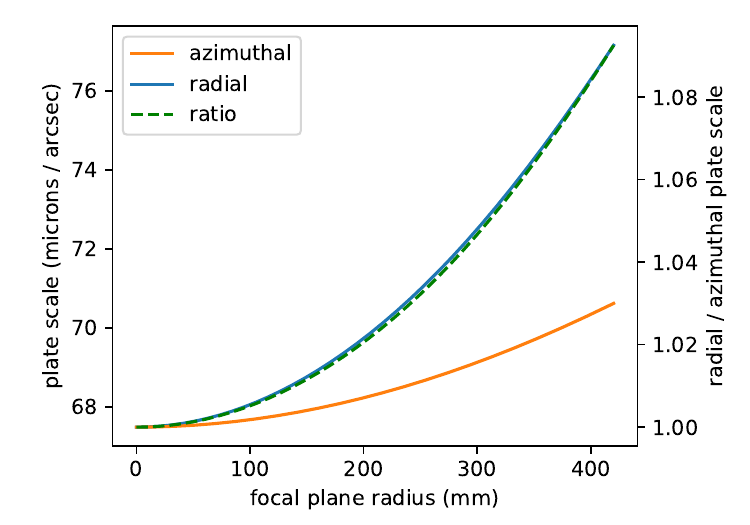}
    \caption{The variation in plate scale in the radial and azimuthal directions as a function of focal plane radius, for DESI.  On the outskirts of the focal plane, fibers see an elliptical area on the sky with an axis ratio of $\sim 1.1:1$.}
    \label{fig:platescale}
\end{figure}

The fiber dither analysis of \textsection\ref{sec:method} is general and can be applied to many multiplex fiber-fed spectrographs.  Most of the detail of the specific spectrograph system comes into the indicator function $A_f$ which specifies what angles on sky are seen by a particular fiber.  For DESI we use the optical modeling to compute the azimuthal (sagittal) and radial (meridional) plate scales at the location of the fiber in the focal plane, as shown in Figure~\ref{fig:platescale}.  We convert the plate scale to an ellipse on the sky using the fixed 107~\mum\ DESI fiber core diameter.  The axis ratio of the ellipse varies with focal plane radius, while the angle of the ellipse varies azimuthally around the focal plane.  We then numerically integrate the point spread function---assumed fixed on the sky---over the appropriate elliptical area for each fiber.  This treatment addresses the fact that the same total fiber offset will lead to different fractional flux losses in different parts of the focal plane.  The plate scale is the only detail of the DESI system design entering the main fiber dither modeling routines.

\subsection{``Lost in space'' dither program}
Between December 2019 and January 2020 while completing the commissioning of the positioners we were unable to successfully place most fibers near enough to target stars to get detectable amounts of flux into the fibers.  Eventually we learned that there was a miscommunication between the hexapod rotation angle used for assigning fibers to locations in the focal plane and the angle used for placing guide cameras on the sky; this led the focal plane to be rotated relative to expectations.  Coupled with slight overall pointing issues and larger than expected distortions in the Fiber View Camera lens, we spent a month unable to get much light into the spectrographs.

During this time we developed an alternative ``lost in space'' dither scheme.  In this scheme we dithered the fibers in a large $5\arcsec\times5\arcsec$ box.  We then moved the telescope through a grid of points with $5\arcsec$ separation so that a few fibers would be on target in each part of the focal plane if the positioning error in any particular region of the focal plane was smaller than 7.5\arcsec.  Studies of which fibers lit up with a given telescope dither, coupled with 0.5\arcsec-scale telescope dithers to get per-fiber offsets for the small number of fibers which were on target eventually identified the aforementioned field rotation issue.  Once this issue was resolved, we were able to place large numbers of fibers on target and start our more detailed dither analysis.

For these ``lost in space'' programs we used ordinary dither target selection, with the ``box'' dither scheme and a large $5\arcsec$ dither scale.  We made a single design and then used telescope dithers to fill out a larger area, aiming to reduce the overall time spent searching for bright stars.

\subsection{Quantities computed for DESI}
\label{subsec:quantities}

The sky-subtracted spectra (\texttt{sframe}) from each of the thirteen tiles in a dither sequence are used as input for the dither analysis. 
 The sky-subtracted spectra are the one-dimensional spectral extractions after flat fielding and sky subtraction \citep{Guy:2023}.  In rare cases, due to fiber collision restrictions, the targets assigned to fibers can vary from exposure to exposure.  We use only spectra from targets where the fiber has been assigned to the same target as the initial, on-target exposure.  We compute the dither amounts $\delta_{e,i}^f$ based on the difference between the on-target location of the star from the initial fiber assignment and the dithered locations of the fibers in subsequent exposures.   For these small dithers we use the flat sky approximation and simply take the difference between the sources coordinates and the intended fiber coordinates in celestial coordinates, adjusting only by $\cos\delta$ in the direction of right ascension.

We first median filter the one-dimensional spectra with an 11-pixel kernel to eliminate any small cosmetic issues in the spectra.  We then compute fluxes by integrating over the spectra in each of the $B$, $R$, and $Z$ cameras, using the wavelength ranges given in Table~\ref{tab:bandpasses}.  We compute the corresponding statistical uncertainties as the square root of the sum of the variances in the fluxes contributing to the integral.  Before entering the dither analysis least squares fitting, these Poisson uncertainties are doubled and an additional five percent is added in quadrature, which we found empirically to provide a better match between the observed flux differences in the dither modeling and the data.

All fluxes have the DESI fiber flat applied.  The fiber flat includes a contribution from the varying throughputs of the different DESI fibers, as well as a factor related to the area of sky seen by each fiber.  We remove the sky area factor using an optical model, since we include this factor separately in the indicator function $A_f$ in the dither modeling (\textsection\ref{sec:method}).

\section{Results for DESI}
\label{sec:results}

The dither analysis demonstrates that the DESI fiber positioning system---the fiber positioner robotics \citep{Silber:2023}, FVC imaging \citep{Baltay:2019}, analysis software \texttt{spotmatch}, and \texttt{PlateMaker} astrometric and metrological system \citep{Kent:2023}---places fibers within about 0.14\arcsec\ of their target locations, or about 0.1\arcsec\ per coordinate.  This accuracy was obtained after including an initial low-order dither-derived correction which we adopted at the end of commissioning; before any dither information was included, the accuracy was around 0.4\arcsec.

\begin{figure*}
    \centering
    \includegraphics[width=\textwidth]{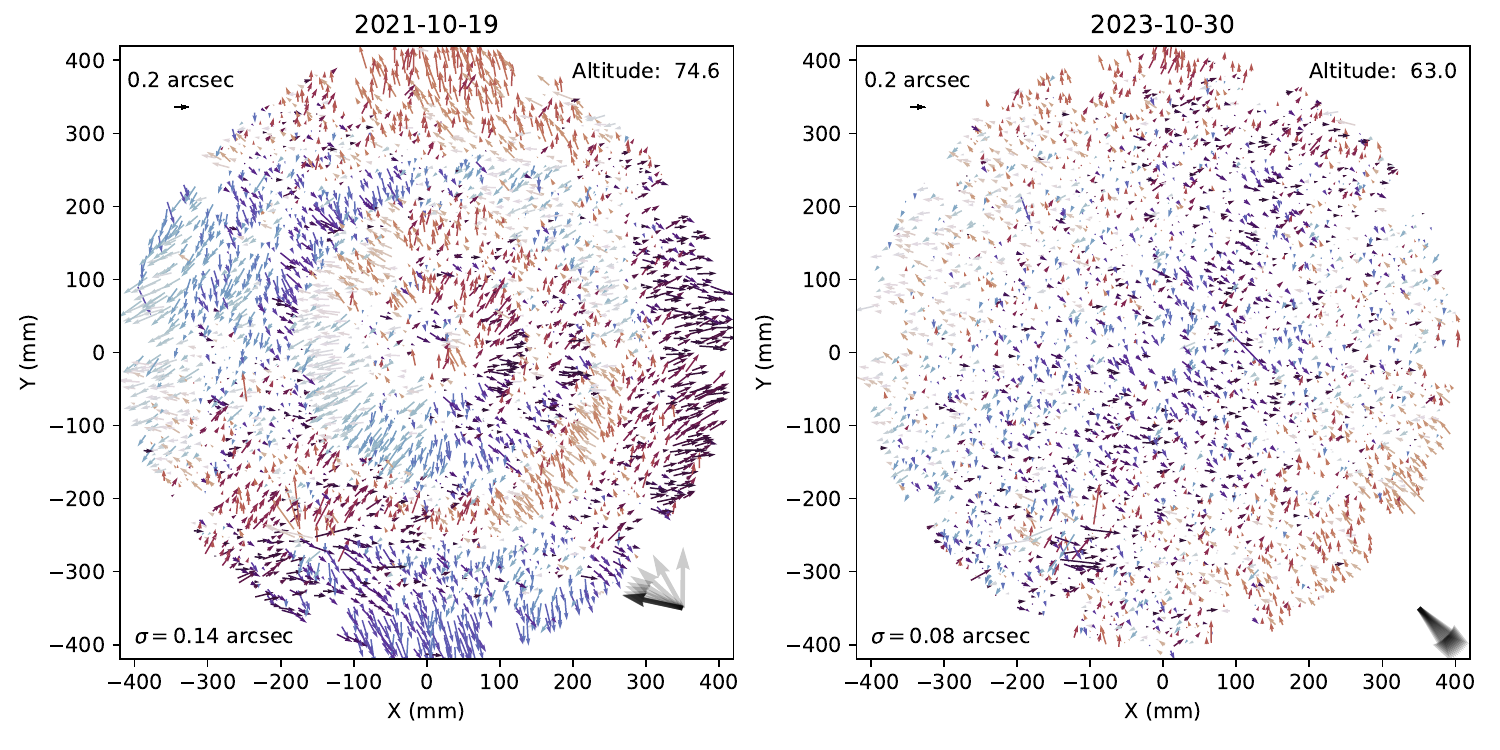}
    \caption{Fiber positioning results from the dither sequences observed on 2021--10--19 and 2023--10--30.  Each arrow shows the derived systematic fiber positioning offset $\delta^f_0$ for the sequence.  Arrows are colored according to their angle, to make coherent regions pointing in the same direction more apparent.  The arrow in the upper left corresponds to 0.2\arcsec.  The root-mean-square offset in the $x$ and $y$ directions is given in the lower left.  The altitude of the observation is given in the upper right, and the directions to zenith for each exposure in the sequence are shown in the lower right.  When multiple exposures have the same direction to zenith the arrow appears darker.  The sequence from 2021--10--19 was observed before the additional higher-order corrections of this work, while the one from 2023--10--30 was observed after those corrections had been incorporated into the pipeline.}
    \label{fig:ditherquiver}
\end{figure*}

This work has enabled a higher-order static distortion map that has further improved positioning.  After inclusion of this higher order map, DESI places fibers within about 0.08\arcsec\ of their target locations, or about 0.06\arcsec\ per coordinate.  This correction has been applied to DESI positioning since 2023--10--30.  Figure~\ref{fig:ditherquiver} illustrates the results for two dither sequences, one before the improvements to positioning, and one after.  Before the new corrections, the fiber positioning offsets have a distinctive pattern across the focal plane suggestive of residual distortions in the fiber view camera lens, mid-scale errors in the polishing of the DESI optics \citep{Miller:2023}, or inhomogeneities in the refractive index of the glass used in the corrector lenses.  After correction, only a very small trend pointing in the direction of zenith remains, potentially suggesting very small systematics in the atmospheric dispersion correction.

\begin{figure*}
    \centering
    \includegraphics[width=\textwidth]{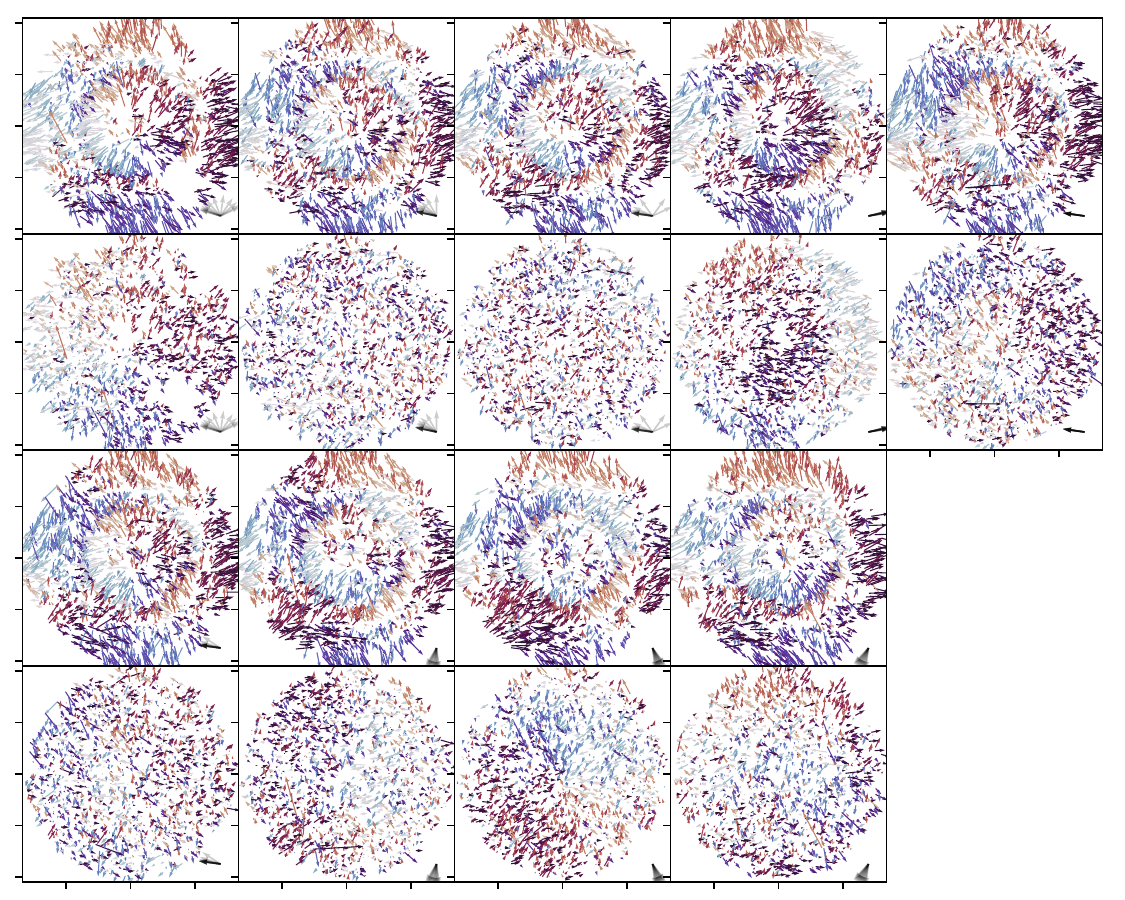}
    \caption{Position offsets (first, third rows) and residuals after correction for a static position error (second, fourth rows), for the nine dither sequences that are the focus of this work.  Colors and axes labels are as in Figure~\ref{fig:ditherquiver}.}
    \label{fig:quivergallery}
\end{figure*}

These offsets are very consistent from dither sequence to dither sequence over 2021--2023.  Figure~\ref{fig:quivergallery} shows the nine sequences before and after removing a static systematic positioning error.  There is good agreement among the sequences, and the static model subtracts off the sequences cleanly.  The most significant residuals are in the high airmass sequence from 2021--10--24, likewise suggesting issues in the atmospheric dispersion correction.

As indicated in Table~\ref{tab:dithersequences}, the dither analysis shows that positioning has a root-mean-square error of about $\approx 0.14\arcsec$ in two dimensions, reducing to $\approx 0.08\arcsec$ error after correcting for a static correction term.  These values are derived from the dither analysis fits to the positions of each individual fiber.  They are a combination of true positioning offsets and uncertainty in the dither analysis, and so represent an upper bound on any systematic in positioning uncertainty that is constant over a dither sequence.

The formal uncertainties in the dither analysis fits are very small, 0.012\arcsec, and do not contribute significantly to the overall residuals.  However, systematic uncertainties in the dither analysis may be much larger.  For example, an error in the PSF shape (assumed constant across the focal plane in the dither analysis) will translate into different errors in the derived fiber positions depending on the set of dithers $\delta^f_{e,i}$ in the sequence for each positioner $i$, leading to the kind of uncorrelated noise seen in the residuals in Figure~\ref{fig:quivergallery}.  On the other hand, we do not see any correlation between $\langle \delta_{e,i}^f \rangle$ and $\delta_{0,i}^f$, which one might expect from an error in the PSF.

Another potential source of positioning error is uncertainty in the centroids in the FVC images, due to photon noise and turbulence in the volume of air between the FVC and the corrector.  These effects contribute roughly 0.05\arcsec\ in typical conditions to the positioning errors \citep{Kent:2023}.  Because photon noise and turbulence vary from exposure to exposure in the hour-long dither sequence, however, we do not expect the dither sequence to be very sensitive to them.  The source of the remaining, spatially uncorrelated $0.08\arcsec$ scatter in the dither-derived positioning error is not well understood, and may be due to mid-scale frequency errors in the corrector optics fabrication.

One interesting feature of Figure~\ref{fig:ditherquiver} is the spot near $(-200~\mathrm{mm}, -250~\mathrm{mm})$ where the dither analysis reports large ($\sim 1\arcsec$) positioning errors.  Rays of light originating in this region of the focal plane pass through a divot on the front surface of corrector C3 on their way to the FVC \citep{Miller:2023}.  This divot covers only a small fraction of the area of C3, but the FVC sees essentially only the chief ray passing through the corrector, and so likewise sees only a tiny fraction of each lens.  As a result, fibers in this region appear fainter than usual and have a peculiar point spread function, leading them to be more challenging to position.  In DESI, we have not attempted to map the distortions caused by this divot in any detail and simply accept worse positioning in this region, which contains roughly 37 fibers.  Note that because the C3 divot only affects a small portion of the full aperture, it does not affect spectroscopy appreciably, and only affects positioning.

\subsection{Chromatic performance}
\label{subsec:chromatic}

\begin{figure}
    \centering
    \includegraphics[width=\columnwidth]{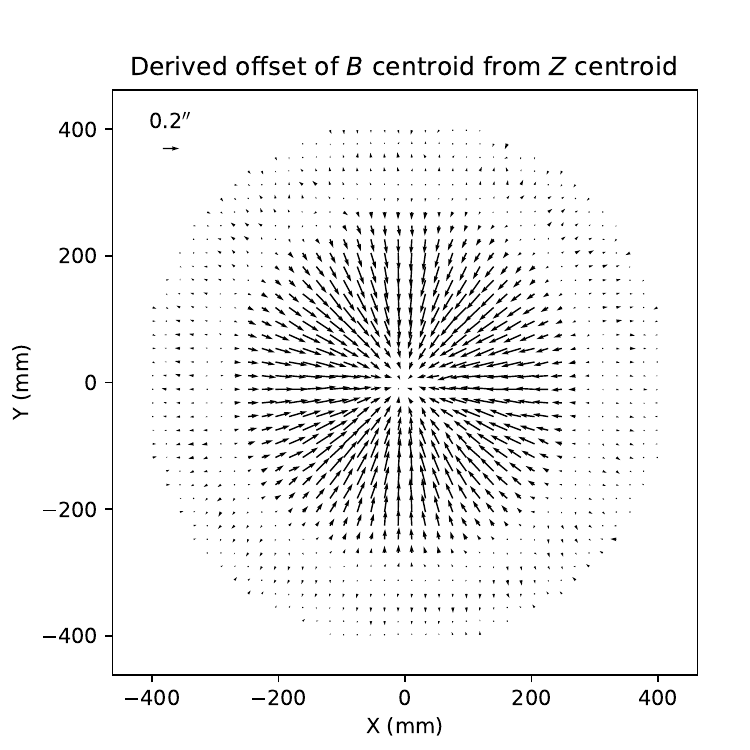}
    \caption{Dither-inferred differences in the centroid of the light reaching the focal plane; vectors point from the centroid of the light inferred for the dither $Z$ band to the center inferred for the dither $B$ band.  The pattern is strongly radial, consistent with expectations from ray trace models.  The arrow in the upper left corresponds to a centroid difference of 0.2\arcsec.}
    \label{fig:chromaticquiver}
\end{figure}

Fiber dither analyses enable tests that light of different wavelengths is being focused together to the same point in the focal plane.  This verifies that the optical system is performing as expected and that the atmospheric dispersion corrector (ADC) is functioning.  Because we are able to separately measure the location of the centroid of the light at different wavelengths in the dither analysis, changes in the centroid of the light with wavelength appear in the differences $\delta_{0,i,\lambda_1}^f - \delta_{0,i, \lambda_2}^f$ for dither analyses performed at different wavelengths $\lambda_1$ and $\lambda_2$.  Figure~\ref{fig:chromaticquiver} shows an example of the derived chromatic offsets between the $B$ and $Z$ bands, using the static dither offset maps derived from all nine dither sequences.

\begin{figure}
    \centering
    \includegraphics[width=\columnwidth]{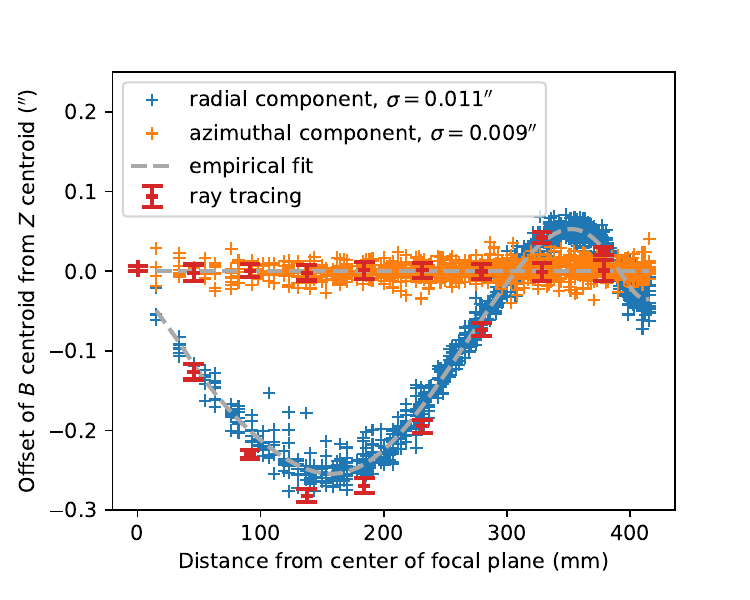}
    \caption{Comparison between dither-derived chromatic offsets and ray-tracing models.  The dither-derived offsets closely match the ray-trace modeling.  The RMS dispersion around an empirical fits to the azimuthal and radial components of the dither-derived offsets is only about 0.01\arcsec.}
    \label{fig:chromaticazrad}
\end{figure}

Offsets between the $B$ and $Z$ bands peak at about $0.25\arcsec$ and are almost entirely radial.  This closely matches expectations from ray tracing, as illustrated in Figure~\ref{fig:chromaticazrad}.  For the ray tracing, we use only centroids for monochromatic light at two wavelengths in the $B$ and $Z$ cameras, averaging those together to match the center of our dither $B$ and $Z$ bands without attempting to match the detailed spectrum of typical DESI targets or the DESI throughput.  We expect that the very small remaining discrepancy between the ray tracing and dither results is attributable to that approximation.  The azimuthal residuals from zero and the radial residuals from a simple empirical polynomial fit are only about $0.01\arcsec$.

For a final application of the chromatic dither analysis, we compare the dither offsets at different wavelengths for a single dither sequence after removing the static model developed over all nine sequences.  This provides another estimate of the uncertainty in a dither-derived position offset; we expect all signal that we know about to cancel out of these differences.  These residual position offsets have root-mean-square differences ranging from 0.021\arcsec\ to 0.028\arcsec\ for the different sequences.  Meanwhile the estimated uncertainties from the dither fits are around 0.021\arcsec\ for each of the fits.  This is decent agreement, but it is important to note that the estimated uncertainties are driven by the empirical systematic flux uncertainties added on top of the Poisson uncertainties in the fluxes (\textsection\ref{subsec:quantities}); there are in any case important systematic uncertainties that we have not yet addressed.  We take this as saying that a signal larger than 0.021\arcsec\ may well be real, which would imply that most of the remaining $\sim 0.08\arcsec$ positioning residuals stem from actual failures to position the fibers correctly.

\subsection{Seeing and throughput measurements}
\label{subsec:seeingthroughput}

\begin{figure*}
    \centering
    \includegraphics[width=\textwidth]{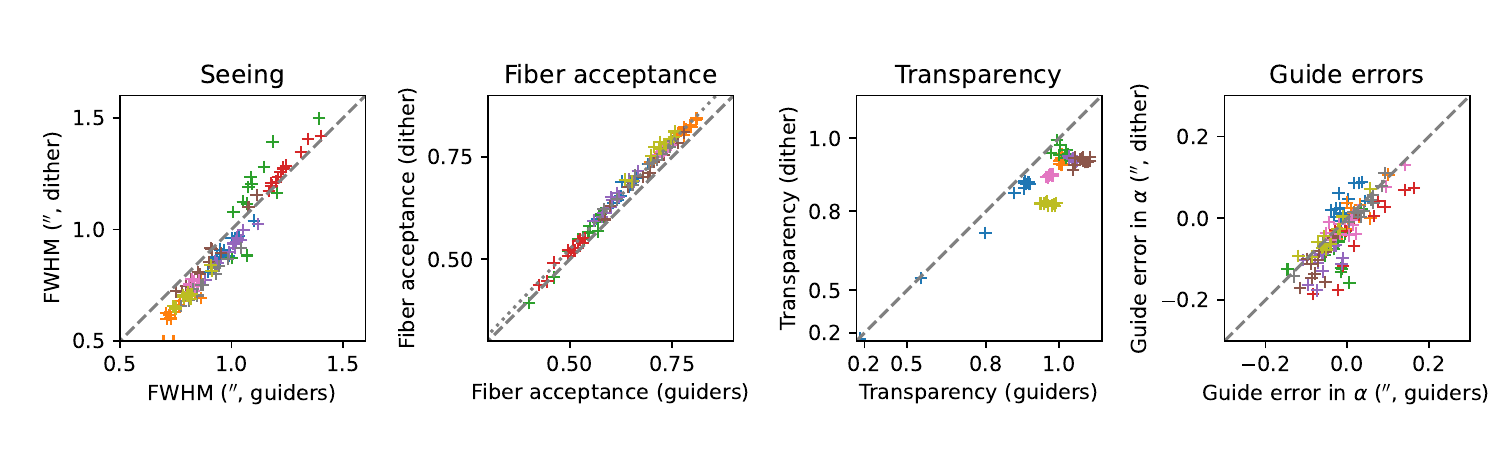}
    \caption{Comparison between dither-derived measurements and guider-derived measurements.  Different dither sequences are shown with different colors.  The first panel compares the FWHM, and the second panel compares the expected amount of light entering the fiber for a well-centered star 288~mm from the center of the focal plane.  The dashed lines show the one-to-one line, while the dotted line shows 1.05-to-1.  The dither data match the guider data well, with an RMS dispersion of 0.13\arcsec\ in FWHM and 0.01 in the fraction of flux entering a fiber.  The third panel and fourth panels show comparisons of the transparency and guiding errors, again as derived from the guiders and dither analysis. 
    \label{fig:gfaseeing}}
\end{figure*}

The dither analysis solves for the shape of the point spread function.  DESI has six guide cameras that can also be used to directly measure the point spread function.  We compare these two different measurements of the point spread function in Figure~\ref{fig:gfaseeing}, and find good agreement.

The GFA and dither-derived PSF full-width-at-half-maxima (FWHM) agree well, with a dispersion of 0.13\arcsec.  There is a modest trend where in very good seeing the dither analysis reports better seeing than the guiders, while in worse seeing the dither analysis reports somewhat worse seeing.  Another way of looking at this trend is to compare the dither-inferred estimates of the amount of flux entering a fiber with what would be inferred from the guide images, as shown in the second panel of Figure~\ref{fig:gfaseeing}.  The dither-derived measurements consistently find roughly 5\% more flux entering a fiber than found in the guide analysis.  The source of this discrepancy is not clear.  This places a 5\% systematic uncertainty on our ability to measure the total DESI system throughput with the dither analysis.

The third panel in Figure~\ref{fig:gfaseeing} compares the dither-derived transparencies with equivalent values from an analysis of the guide cameras.  Both sets of values are corrected for airmass to a reference airmass of 1.  The dither-derived values are ``absolute'' in the sense that they compare the number of photons observed in the spectrograph to expectations from the dither-derived fiber-acceptance fraction and the expected total throughput of the system, and attribute any overall difference to atmospheric transparency.  Meanwhile the guider-derived values compare the observed zero point to a reference zero point derived on a photometric night.  The dither-derived transparencies are consistently lower than the guider-derived transparencies, with a median difference of 7\%.  Most of that is explained by the 5\% difference in derived fiber acceptance fraction seen in the second panel of Figure~\ref{fig:gfaseeing}.  The scatter in the difference between the two measures of transparency is 4\%.  Artificially correcting the dither transparencies for the differences between the dither and guider fiber acceptance fractions reduces the median difference to 2.5\% and the clipped scatter to 2.8\%, but it remains the case that the dither analysis measures systematically different transparencies than expected from the guider analysis by up to 9\%.  It is not known what drives the remaining scatter in this quantity.  On the other hand, the derived throughput matches expectations at the 3\% level after accounting for the difference between the guider- and dither-inferred seeing, even if we do not yet know what drives those different seeing measurements.

Finally, the fourth panel of Figure~\ref{fig:gfaseeing} compares the telescope offsets derived from the dither analysis with those derived from analysis of the guider images.  The guide cameras are used to measure and correct any drift of the telescope from the target location on the sky, but latency and noise in the process mean that guider-measured telescope pointing offsets are not immediately corrected.  We here take the mean guider offset over the whole exposure as a measure of the guider-estimated telescope offset.  These offsets correlate well with the equivalent dither-measured quantities, with no significant mean offset in right ascension and a scatter of about 0.04\arcsec.  On the one hand, this is good agreement: 0.04\arcsec\ corresponds to less than 3~\mum\ in the focal plane.  On the other hand, given that we can measure chromatic positioner offsets to better than 0.03\arcsec, it is surprising that telescope offsets, which can be constrained with all $\sim 4,000$ stars on each exposure, are not better measured; it seems likely that the guiders will be more accurate than the dither analysis here.  The comparison is complicated by the fact that overhead in the guiders means that they spend only roughly half of the exposure time collecting photons, so some discrepancy between the guiders and the dither analysis may be expected.  Finally, given the subtlety in defining the center of a PSF and the simple PSF modeling taking place in the dither analysis, we consider this level of agreement good.

\subsection{Flux comparisons}
\label{subsec:flux}

\begin{figure}
    \centering
    \includegraphics[width=\columnwidth]{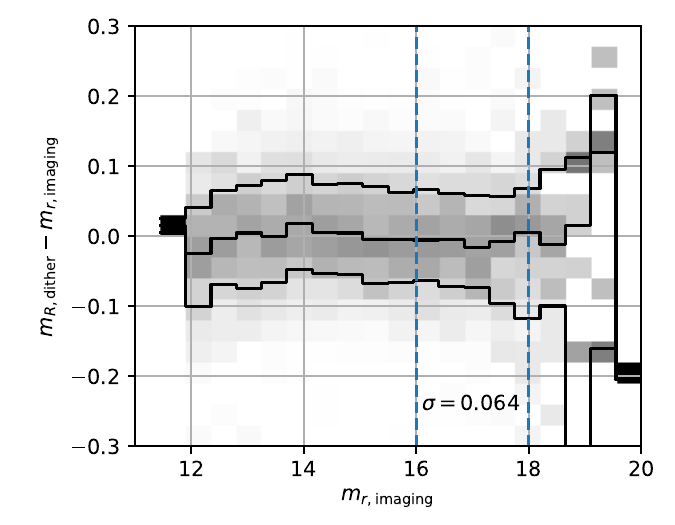}
    \caption{Comparison between dither-derived and imaging $r$ fluxes for the dither sequence taken on 2023--10--19.  The distribution of differences between the dither-derived magnitudes and imaging magnitudes is shown as a function of the imaging magnitude, in the $r$ band.  Contours show the 16th, 50th, and 84th percentiles in each bin of imaging magnitude.  There is good agreement all the way from 12th mag to 20th mag, The RMS scatter $\sigma$ in the difference over $16 < r < 18$ is 0.064 mag.
    \label{fig:fluxcompare}}
\end{figure}

The last set of parameters measured from the dither analysis is the total fluxes of the individual stars.  These are adjusted through the ``transparency'' parameter for each exposure  to match expectations from the DESI Legacy Imaging Surveys \citep{Dey:2019, Zou:2017}, but the scatter provides another test of the accuracy of the dither analysis.  Figure~\ref{fig:fluxcompare} shows the imaging fluxes as compared with the dither-derived stellar fluxes, for data taken as part of the 2023--10--19 dither sequence.

The agreement between the dither fluxes and imaging fluxes is acceptable, with an RMS scatter of $0.064$ mag.  However, formal uncertainties are roughly only 0.02~mag, and we do not know the source of the additional uncertainty.  Effects like variation in the PSF over the focal plane or errors in the color transformation seem plausible but are largely ruled out by the absence of noticeable trends in residual versus location in the focal plane or imaging color.

\section{Conclusions \& Future Work}
\label{sec:conclusion}

Most astronomical spectrographs rely on the ability to precisely focus light from target sources onto slits or fibers in order to efficiently measure properties of sources.  We have presented a technique for measuring systematic errors in the positioning of fibers for large, fiber-fed, multi-object spectrographs.  In our approach, fibers are dithered around target stars following known patterns over a series of exposures; these dithers are different for each targeted star in each of a series of exposures.  By measuring the amount of light that enters the fiber as a function of the known dither, we are able to solve for any systematic fiber offset over the dither sequence, in addition to the PSF delivered to the focal plane, telescope guiding errors, and the sky transparency or system throughput.  We apply this technique to DESI and demonstrate that DESI can position fibers with an accuracy of about 0.08\arcsec.  Comparison between measured fiber offsets at different wavelengths suggest that we may obtain accuracies as good as 0.02\arcsec, similar to the estimated statistical uncertainties, but it is unclear if some source of systematic uncertainty is canceling in those comparisons.  Comparison with DESI GFA images likewise indicates that we are able to accurately measure the seeing, transparency, and guide errors that DESI sees, providing strong evidence that our measured fiber offsets are reliable.  This technique can be applied to any multi-object spectrograph to robustly measure positioning errors of each fiber, and we expect that other instruments like SDSS-V \citep{Kollmeier:2017} and the Prime Focus Spectrograph \citep{Takada:2014} will also adopt it.

Correctly positioning fibers brings important benefits.  It maximizes the light down fibers, speeding measurements.  For cosmology-focused instruments like DESI, it brings the additional benefit in delivering a more homogeneous survey.  While improving fiber positioning from 0.14\arcsec\ offsets to 0.07\arcsec\ offsets ``only'' improves the survey speed by roughly 3\%, it also reduces the variation in the redshift success rates among the different DESI fibers, simplifying the modeling needed to transform DESI from a list of redshifts into cosmological parameters.

We expect that future work will improve upon our results here in measuring the total throughput of spectrograph systems.  We were only able to reproduce transparencies measured by the GFAs at the 4\% level, preventing us from making confident statements about the DESI system throughput to better than that precision.  Still, these measurements allow an important cross-check on other measurements of total system throughput.

Because we are able to measure the PSF delivered to the focal plane using fiber dithering, in principle we can use this technique to measure the $z$ height of each DESI fiber tip, or equivalently the $z$ offset needed to bring each fiber into perfect focus.  One approach for achieving this would be to do a dither sequence on an intentionally out of focus image, and deriving the shape of the out-of-focus donut of light delivered to each fiber by dithering around it.  We have taken only initial steps in this direction in DESI so far, but future work should be able to make measurements of this kind.

Fiber dither sequences exercise much of the DESI system.  Guiding, focusing, positioning, spectroscopic throughput, and the spectroscopic pipeline are all stringently tested.  Many components needed to come together to enable the $0.1\arcsec$ positioning presented here, but of course this positioning is only one small step toward's DESI's goal of making the world's largest three-dimensional map of the universe.  We are looking forward to measuring millions of spectra positioned at very precise locations on the sky over the coming years as DESI executes its main survey.

\acknowledgements

This material is based upon work supported by the U.S. Department of Energy (DOE), Office of Science, Office of High-Energy Physics, under Contract No. DE–AC02–05CH11231, and by the National Energy Research Scientific Computing Center, a DOE Office of Science User Facility under the same contract. Additional support for DESI was provided by the U.S. National Science Foundation (NSF), Division of Astronomical Sciences under Contract No. AST-0950945 to the NSF’s National Optical-Infrared Astronomy Research Laboratory; the Science and Technology Facilities Council of the United Kingdom; the Gordon and Betty Moore Foundation; the Heising-Simons Foundation; the French Alternative Energies and Atomic Energy Commission (CEA); the National Council of Humanities, Science and Technology of Mexico (CONAHCYT); the Ministry of Science and Innovation of Spain (MICINN), and by the DESI Member Institutions: \url{https://www.desi.lbl.gov/collaborating-institutions}. Any opinions, findings, and conclusions or recommendations expressed in this material are those of the author(s) and do not necessarily reflect the views of the U. S. National Science Foundation, the U. S. Department of Energy, or any of the listed funding agencies.

The authors are honored to be permitted to conduct scientific research on Iolkam Du’ag (Kitt Peak), a mountain with particular significance to the Tohono O’odham Nation.

\vspace{5mm}
\facilities{DESI}
\software{astropy \citep{astropy2013a, astropy:2018, astropy:2022}}

\bibliography{fiberdither}
\end{document}